\documentclass[%
 reprint,
 amsmath,amssymb,
 aps,
pra,superscriptaddress,
]{revtex4-2}

\usepackage{graphicx}
\usepackage{dcolumn}
\usepackage{bm}

\usepackage[dvipsnames]{xcolor}

\begin{document}

\preprint{APS/123-QED}

\title{Memory-enhanced quantum extreme learning machines for characterizing non-Markovian dynamics}

\author{Hajar Assil}
\email{assil.hajar@etu.uae.ac.ma}
\affiliation{Laboratory of R\&D in Engineering Sciences, Faculty of Sciences and Techniques Al-Hoceima, Abdelmalek Essaadi University, Tetouan, Morocco}

\author{Abderrahim El Allati}
\affiliation{Laboratory of R\&D in Engineering Sciences, Faculty of Sciences and Techniques Al-Hoceima, Abdelmalek Essaadi University, Tetouan, Morocco}
\affiliation{Université Grenoble Alpes, CNRS, LPMMC, 38000 Grenoble, France.}

\author{Gian Luca Giorgi}%
\email{gianluca@ifisc.uib-csic.es}
\affiliation{%
 Institute for Cross-Disciplinary Physics and Complex Systems (IFISC) UIB-CSIC, Campus Universitat Illes Balears, 07122 Palma de Mallorca, Spain.
}%

\date{\today}

\begin{abstract}

We use a Quantum Extreme Learning Machine (QELM) for characterizing and estimating parameters of quantum dynamics generated by a tunable collision model. The input to the learning protocol consists of quantum states produced by successive system–environment interactions, while the reservoir is implemented as a disordered many-body quantum system evolving under a fixed Hamiltonian. We systematically explore how extending the QELM feature space, through the inclusion of temporal information and additional observables, affects estimation performance. Our results demonstrate that temporal extensions of the feature vector consistently and significantly enhance estimation accuracy relative to the baseline protocol. Notably, incorporating memory from earlier time steps yields the most substantial and robust improvements, whereas extensions based solely on additional observables offer only marginal gains. Crucially, the advantage conferred by temporal memory becomes increasingly pronounced as the dynamics become more strongly non-Markovian, indicating that environmental memory effects serve as a constructive resource for learning. 
\end{abstract}

\maketitle


\section{\label{sec:level1} Introduction}

Quantum information processing relies heavily on our ability to accurately characterize the dynamics of quantum systems as they evolve in the presence of environmental interactions. The resulting quantum channels, mathematical descriptions of how quantum states transform, are central to understanding and controlling open quantum systems in contexts ranging from quantum computing and communication to sensing and metrology  \cite{breuer2002theory, rivas2012open, liu2011experimental, chruscinski2022dynamical, de2017dynamics}. However, reconstructing these channels is often challenging, especially when the underlying dynamics involve memory effects, temporal correlations, or complex system–environment couplings.


In light of these challenges, the development of scalable and resource-efficient strategies to estimate and characterize quantum states and processes has become a central objective in quantum information science, with broad implications for quantum metrology and emerging quantum technologies \cite{PhysRevLett.109.233601,Gebhart_2023, polino2020photonic, czerwinski2022selected,ma2025machine}. Across these lines, numerous studies have shown that integrating machine learning techniques into quantum state estimation frameworks can substantially improve their efficiency and scalability \cite{Suprano_2024, Cerezo_2022, Giordani_2020, cimini2023deep,lohani2020machine,ma2025machine, Gao_2018, Ghosh_2019}. In this context, Quantum Extreme Learning Machines (QELMs) have emerged as a promising framework for quantum estimation tasks \cite{PhysRevA.111.022412, Xiong_2025, Vetrano_2025, gili2026learning}. Building on the classical extreme learning machine paradigm, a QELM employs a fixed-structure quantum dynamical system to perform a static, one-shot transformation of input data into a high-dimensional Hilbert space, generating rich nonlinear feature representations. This strategy is particularly well-suited to supervised tasks such as classification and state estimation, where temporal dynamics are not essential. By capitalizing on the exponential growth of degrees of freedom in quantum systems while restricting training to a simple linear readout layer, QELM enables efficient inference with reduced optimization complexity.\cite{Innocenti_2023, Mujal_2021, De_Lorenzis_2025, Vetrano_2025, PhysRevA.111.022412, PhysRevA.107.042402}. These considerations make QELM suited for learning and characterizing parametrized open quantum system dynamics.

In this work, we propose and analyze a QELM-based framework for the reconstruction and parameter estimation of quantum channels generated by a tunable collision model. Collision models provide a flexible and analytically tractable way to simulate open system dynamics, allowing precise control over parameters such as system–environment coupling and environmental memory \cite{Campbell_2021, ciccarello2022quantum,ziman2004descriptionquantumdynamicsopen, PhysRevA.72.022110,PhysRevLett.126.130403,Saha_2024, Corretal2025}. Using such a model to produce sequences of output states, we generate quantum data that encode the structure of the channel. These data are then processed by a QELM, whose reservoir is realized by a many-body quantum system evolving under a fixed Hamiltonian.\\

Our primary objective is channel discrimination and parameter estimation—identifying whether a given channel is Markovian or non-Markovian, and inferring key parameters such as the coupling strength and depolarization rate. We systematically investigate how extending the QELM feature space with temporal information and additional observables affects reconstruction accuracy, quantified by the normalized mean squared error (NMSE). Our results demonstrate that incorporating memory from earlier time steps significantly improves channel characterization, outperforming approaches based solely on additional observables. Notably, these enhancements are most pronounced in regimes where the channel exhibits strong memory effects, underscoring the role of temporal correlations in enabling accurate reconstruction.\\

The manuscript is structured as follows. Section \ref{sec:model} introduces the collision model used to generate tunable non-Markovian quantum dynamics. Section \ref{sec:protocol} QELM protocol and our extensions for incorporating temporal memory. Section \ref{sec:results} presents the results, analyzing parameter estimation for both the coupling strength and the depolarization rate, and demonstrates the performance enhancement from memory-augmented feature vectors. Finally, Section V summarizes our conclusions and outlines potential future directions.

\section{Model System} \label{sec:model}
We employ a non-Markovian collision model based on
the framework introduced in  \cite{Rijavec_2025}. The collision model adopted is used to investigate the non-Markovian dynamics of a system of qubits interacting with an environment. In contrast to standard approaches, where memory effects are often introduced through structured reservoirs, the scheme controls the degree of non-Markovianity via two mechanisms: the system
interacts locally with the same reservoir "bath" qubits and applies a depolarizing channel to the bath \cite{Rijavec_2025}. {In Ref. \cite{sannia2025non}, the channel introduced in \cite{Rijavec_2025} was employed as a quantum reservoir, due to the tunability of its non-Markovian character. In this work, we adopt a different approach: we use the collision model to generate streams of quantum data, which is then fed into a separate many-body reservoir (a transverse-field Ising model), which acts as a QELM. 

\textit{System -}
We consider an isotropic one-dimensional chain composed of
$N$ qubits, which we denote as the ``\textit{system qubits}'' $S_n$ with $n=1,\dots,N$. The dynamics of this chain is governed by a Heisenberg Hamiltonian:
\begin{equation}
    \hat{H}_{{ \rm syst }}=\frac{1}{2} \sum_{n=1}^N J_{{ \rm syst }}^n\left(\hat{X}_n \hat{X}_{n+1}+\hat{Y}_n \hat{Y}_{n+1}+\hat{Z}_n \hat{Z}_{n+1}\right) ,
     \label{Hsyst}
\end{equation}

where $J_{\text{syst}}^n \in [-J_s, J_s]$ are coupling constants and $\hat{X}_n, \hat{Y}_n, \hat{Z}_n$ denote the standard Pauli operators acting on the $n$-th site.

\textit{Bath-}
The environment is modeled as a reservoir of $N$ interacting "\textit{bath qubits}" $Q_m (m= 1, \ldots, N)$  described by a
similar Heisenberg Hamiltonian:

\begin{equation}
     \hat{H}_{\text {bath}}=\frac{1}{2} \sum_{m=1}^N J_{\text {bath}}^m\left(\hat{\mathcal{X}}_m \hat{\mathcal{X}}_{m+1}+\hat{\mathcal{Y}}_m \hat{\mathcal{Y}}_{m+1}+\hat{\mathcal{Z}}_m \hat{\mathcal{Z}}_{m+1}\right)  
     \label{Hbath}
\end{equation}

with coupling constant $J_{\text {bath}}^m \in  [-J_s,J_s]$ and $J_{s} \in \mathbb{R}$ $\forall$  $m=1, \ldots, N$ and $\hat{\mathcal{X}}_m$, $\hat{\mathcal{Y}}_m, \hat{\mathcal{Z}}_m$ the Pauli operators of the m-th bath qubit.

The non-Markovian dynamics are generated through a four-step process that governs the interaction between the system and the bath at each iteration, as illustrated in Fig.~\ref{fig:placeholder}. In each collision step:
\begin{enumerate}
    \item \textbf{Exchange phase}: Each system qubit interacts with its nearest bath qubit through a partial swap operation: 
    \begin{equation}
        \hat{P}_m(\chi)=\cos (\chi) \mathbb{I}+i \sin (\chi) \hat{\mathcal{S}}_m
        \label{pswap}
    \end{equation}
    where $\hat{\mathcal{S}}_m$ is the swap operator between the m-th system and bath qubits, and $\chi \in [0, \pi / 2]$ is a parameter that regulates the strength of the interaction.
    \item \textbf{Depolarization phase}: 
    After the interaction, a depolarizing channel acts on the bath qubits.
    \begin{equation}
        \Delta_{\lambda}^m\left(\rho \right)=\sum_{i=0}^3 \hat{K}_i^m \rho  \hat{K}_i^{m \dagger},
    \end{equation}
   with Kraus operators $\hat{K}_0^m = \sqrt{1-3\lambda/4} \mathbb{I}_m$ and $\hat{K}_{1,2,3}^m = \sqrt{\lambda/4} \hat{\sigma}_{x,y,z}$., and $0 \leq \lambda \leq 1$.

    In this step, we can control how much information is lost to the environment and therefore control the degree of non-Markovianity of the environment.\\
    Importantly, $\lambda$ represents the strength of the depolarizing channels and serves as a control parameter for tuning the level of non-Markovianity. At $\lambda = 1$, the bath is reset to a completely mixed state at each step, yielding purely Markovian dynamics. Conversely, $\lambda = 0$ preserves bath correlations, maximizing memory effects.
    
    \item \textbf{Bath evolution}:
    In this stage, the reservoir qubits evolve under their Hamiltonian eq.~\ref{Hbath}, which allows information to spread within the environment.

    \item \textbf{System evolution}: 
    The system qubits evolve under their Hamiltonian eq.~\ref{Hsyst} for the same time interval $\Delta t$. 
    
\end{enumerate}
\begin{figure}
    \includegraphics[width=0.8\linewidth]{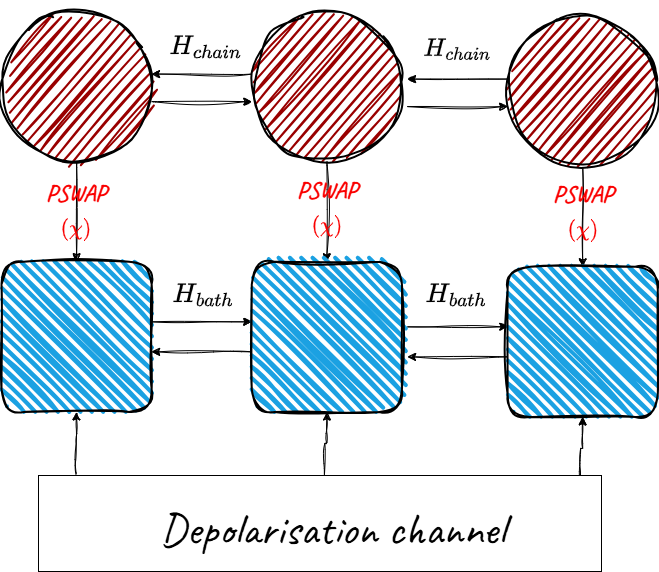}
    \caption{Schematic of the collision model with three system qubits $S_i$ (red) and three reservoir qubits $Q_i$ (blue). 
    The protocol evolves in successive stages: each system qubit first exchanges information with its corresponding reservoir qubit via a 
    PSWAP operation of strength $\chi$. 
    The reservoir qubits are then subject to a depolarizing channel before undergoing unitary evolution governed by $H_{\text{bath}}$. 
    Finally, the system qubits evolve under their internal interactions described by $H_{\text{chain}}$. }
    \label{fig:placeholder}
\end{figure}

\section{QELM Protocol}\label{sec:protocol}
Quantum reservoir computing (QRC) and QELM are two closely related paradigms that make use of the dynamics of complex quantum systems as computational resources \cite{PhysRevApplied.8.024030,Ghosh_2019,Mujal_2021,Nokkala2021}. Both approaches exploit the natural evolution of a fixed, typically disordered, many-body quantum system to nonlinearly map input states into a high-dimensional feature space. A simple linear readout is then trained on these features to perform tasks such as classification, state estimation, or parameter prediction. The key distinction lies in their original motivation: QRC traditionally emphasizes the temporal processing capability of the reservoir, using its intrinsic memory to process time-series data, while QELM focuses on the instantaneous mapping of input states, treating the reservoir as a static feature map optimized for one-shot transformations. In practice, however, both frameworks share the same core mechanism, that is, a fixed quantum dynamical system followed by a trained linear output layer. In this work, we adopt the QELM perspective because our primary goal is to estimate the parameters of the collision model ($\chi$ and $\lambda$) from the sequence of output states produced by the system. Nevertheless, as we will show, incorporating temporal information by augmenting the feature vector with past reservoir states significantly enhances estimation accuracy. This effectively bridges the two viewpoints, demonstrating that even within a static estimation task, the memory of the reservoir can be harnessed as a resource.

We now present the QELM protocol used in our work. A QELM is typically composed of three main components:
\begin{enumerate}
    \item \textit{Inputs.} In our case, the inputs to the QELM are the quantum states generated by the collision model protocol. After each collision step, we extract the reduced density matrix of the system, denoted as $\rho_S^{(k)}$. These states are then used as inputs to the reservoir.

    \item \textit{Reservoir.} The reservoir is modeled by a transverse Ising system, whose Hamiltonian reads \cite{PhysRevApplied.8.024030,PhysRevLett.127.100502}: 
    \begin{equation}
        H=\sum_{i<j}^N J_{i j} \sigma_i^x \sigma_j^x+h \sum_{i=1}^N \sigma_i^z,
    \end{equation}
    
    where $i$ and $j$ label the sites of the network; $\sigma_i^a(a=x, y, z)$ are the Pauli matrices acting on the $i-th$ site; $h$ is the value of the external magnetic field; $J_{i j}$ is the spin-spin coupling, randomly selected from a uniform distribution in the range $\left\{-J_s / 2, J_s / 2\right\}$ with $J_s=1$; and $N$ is the number of qubits.
    
    The interaction between the input and the reservoir induces a nonlinear transformation. The evolved reservoir state is given by
    \begin{equation}
        \rho_k(t) = U(H,t)\, \rho_{S}^{(k)} U^\dagger(H,t),
    \end{equation}
    where $U(H,t) = e^{-iHt}$ is the time evolution operator and $\rho_{S}^{(k)}$ is the initial reservoir state.\\
    To extract information from the evolved state, we perform measurements defined by the Pauli operator
     $\sigma_i^z$ acting on each $i-th$ qubit. The expectation values
    $$
    x_i(t)=\left\langle \sigma_i^z\right\rangle=\operatorname{Tr}\left[\sigma_i^z \rho_k(t)\right]
    $$ 
    form the components of the reservoir feature vector.
    
    Thus, each input state $\rho_S^{(k)}$ is mapped by the reservoir dynamics into a high-dimensional feature vector
    \begin{equation}
       \mathbf{x}_k=\left(x_1^k, x_2^k, \ldots, x_N^k\right)^{\top},
       \label{x_k}
    \end{equation}
    
    where the nonlinear transformation arises naturally from the many-body interactions governed by the Hamiltonian. These features are then transmitted to the output layer for linear regression\cite{PhysRevApplied.8.024030}.

    To enhance the expressive power of the protocol, we incorporate temporal memory by augmenting the feature vector at the current time step $k$. As illustrated in Eq.~\eqref{x_k'}, this is achieved by concatenating the instantaneous features $\mathbf{x}_k$ with the reservoir feature vector from an earlier time step $k'$:
     \begin{equation} \hat{\mathbf{x}}_k = \left(x_1^k, \ldots, x_N^k, x_1^{k'}, \ldots, x_N^{k'}\right)^{\top}, \label{x_k'}
    \end{equation}

     This temporal augmentation provides the readout layer with direct access to the dynamical history of the reservoir, effectively allowing it to compare the current system state with a past snapshot. To benchmark the value of this historical information against simply having more instantaneous information, we will compare its performance against a different type of feature space extension. In this alternative approach, we augment the feature vector not with past states, but with measurements from an additional observable at the same time step $k$, specifically replacing the $x_i^{k'}$ components with $\langle \sigma_x^i \rangle^k$. This allows us to directly assess whether the benefits we observe stem from accessing temporal correlations or merely from increasing the sheer amount of data per time step.
    
    \item \textit{Output.} The readout layer performs a linear regression on the extended input features. The predicted output for the $k$-th input is given by
\begin{equation}
    \mathbf{y}_k = W \, \mathbf{x}_k ,
\end{equation}
where $W$ is the trainable weight matrix. These weights are optimized classically by minimizing the squared error between predicted and target values. In matrix form, the optimal solution is given by
\begin{equation}
    W = Y X^\top (X X^\top)^{-1},
\end{equation}
where $X$ is the matrix collecting all extended input vectors and $Y$ contains the corresponding target values (the true parameters $\lambda$ or $\chi$). The optimal weights are found using the Moore-Penrose pseudo-inverse method.

\end{enumerate}

\section{Results and Discussion}\label{sec:results}
    In this work,  we address the problem of quantum channel discrimination by combining a collision model description of open system dynamics with a QELM framework. The primary objective is to determine whether a given channel exhibits predominantly Markovian or non-Markovian behavior. To generate the system dynamics, we consider a collision model consisting of two system qubits interacting sequentially with two reservoir qubits. The interactions are implemented via partial-swap (PSWAP) operations (see Eq.~\ref{pswap}), which allow us to simulate different degrees of information exchange between the system and its environment. By controlling whether the reservoir qubits interact or remain independent, we can switch between Markovian and non-Markovian dynamics, thereby creating a flexible framework for testing channel discrimination. 

The states generated by the collision model are subsequently processed by a QELM consisting of a five-qubit reservoir. As described in the previous section, at each time step the reservoir's output is used to construct a feature vector $\mathbf{x}_k$ (see Eq.~\ref{x_k}). These vectors are then processed by a linear readout, trained to estimate the channel parameters $\lambda$ and $\chi$. The complete protocol, from the collision model dynamics to the QELM classification, is illustrated in the schematic in Fig.~\ref{scheme_protocol}.

\begin{figure}
    \centering
    \includegraphics[width=1\linewidth]{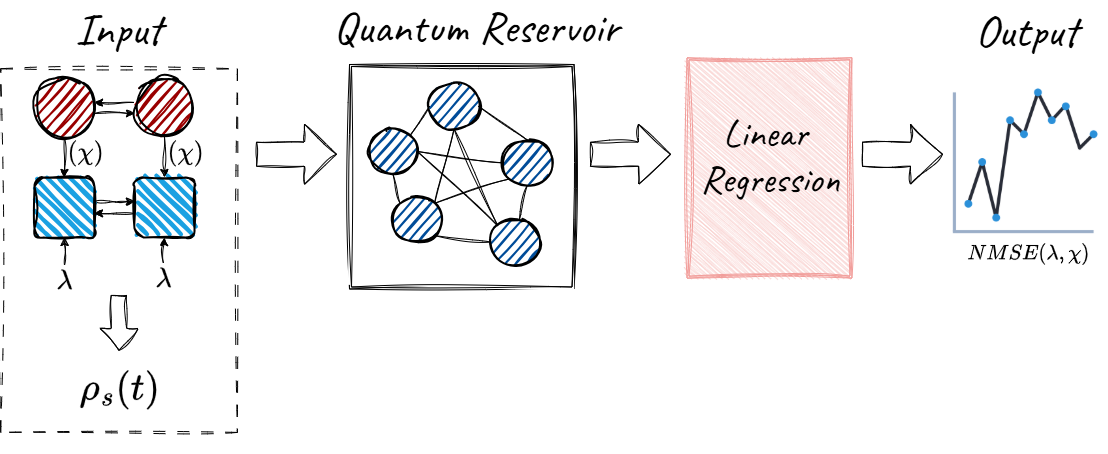}
    \caption{Schematic representation of the protocol. The input states $\rho_s(t)$ are processed by a quantum reservoir, and the resulting features are evaluated by a linear regression readout. 
    The output performance is quantified by the NMSE.}
    \label{scheme_protocol}
\end{figure}

In order to assess the performance of the proposed protocol, we organize the results into two main parts. First, we examine parameter estimation, where the task is to infer either $\chi$ or $\lambda$ while keeping the other fixed. Second, we explore memory-enhanced learning, where we investigate how incorporating the reservoir’s memory improves the QELM’s ability to extract information from the system's dynamics.

\subsection{Parameter estimation}

\begin{figure}[ht]
    \centering
    \includegraphics[width=1\linewidth]{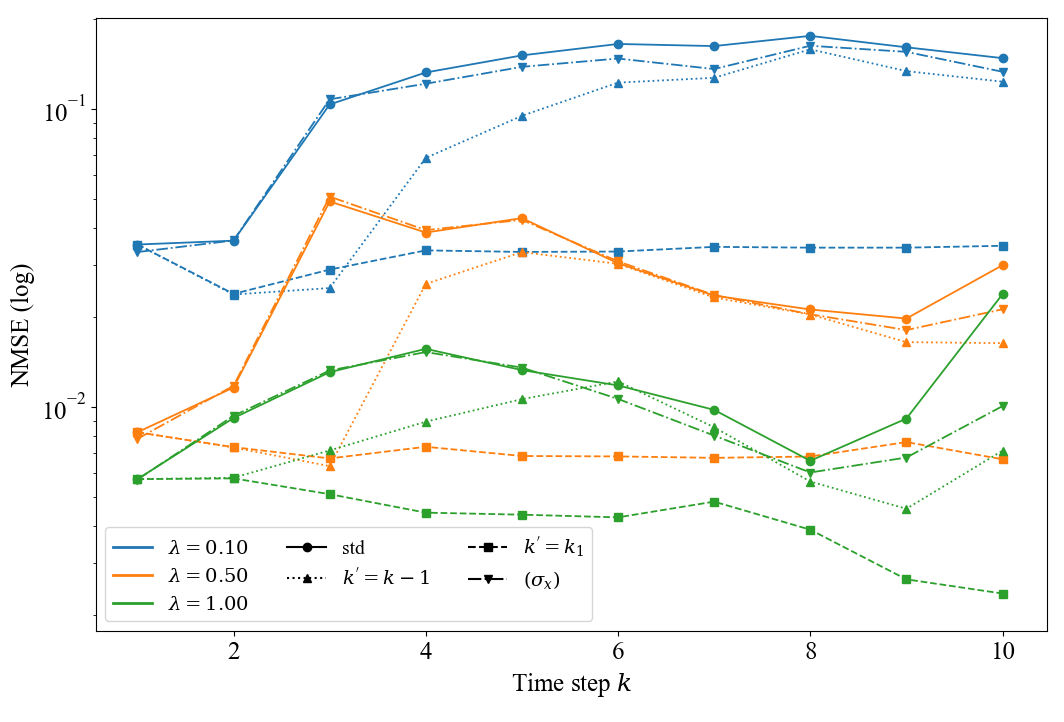}
     \caption{NMSE as a function of the collision step for different QELM feature extensions and coupling strengths $\lambda = 0.10,\; 0.50,\; 1.00$. Solid lines correspond to the baseline QELM without extensions. Dotted lines show the inclusion of the immediate past reservoir state, dashed lines represent the use of a fixed earlier memory step $k_1$, and dash-dotted lines indicate the extension based on the additional observable $\langle\sigma_x\rangle$ (Average of 2000 realizations).}
    \label{fig:omega_ext}
\end{figure}

\textit{Parameter estimation for $\chi$.} In this case, we fixed the parameter $\lambda$ while the interaction strength $\chi$ was varied as a function of the collision time step $k$.
This analysis is conducted across three distinct environmental regimes defined by the depolarization parameter $\lambda$: a highly non-Markovian regime ($\lambda = 0.10$, blue), an intermediate regime ($\lambda = 0.50$, orange), and the purely Markovian limit ($\lambda = 1.00$, green). The precision of the estimation is quantified by the NMSE on a logarithmic scale, as illustrated in Fig.~\ref{fig:omega_ext}.

Our results indicate that the degree of environmental non-Markovianity significantly influences the learning task. The purely Markovian regime ($\lambda = 1.00$) consistently achieves the lowest NMSE across all feature configurations. In contrast, as the dynamics become increasingly non-Markovian (decreasing $\lambda$), the NMSE rises, suggesting that the accumulation of memory effects initially poses a challenge for isolating the coupling parameter $\chi$. 
A central finding of this work is that extending the feature vector to include temporal information is always beneficial for the protocol's performance. While the $(k^{'} = k-1)$ configuration (dotted lines with triangles) provides some improvement, the most significant and robust results are obtained using the $(k^{'} = 1)$ configuration (dashed lines with squares). 

By incorporating the first time step as a reference point, the QELM successfully harnesses temporal correlations as a resource to overcome environmental interference. This memory-augmented approach allows the system to achieve a steady reduction in NMSE as time progresses, eventually outperforming the initial accuracy in the intermediate regime. Furthermore, while measurements in the $\sigma_x$ basis (dot-dashed lines) offer alternative feature sets, they generally do not match the stability and high fidelity of the $(k^{'} = 1)$ memory-enhanced configuration. 
This finding is key: providing the QELM with a direct temporal reference point (the initial state) is significantly more effective than simply providing it with a more detailed snapshot of the current moment (via the $\sigma_x$ measurement). The temporal correlation across time steps encodes information about the system-environment memory that a single-time, multi-observable measurement cannot capture.
Collectively, these results demonstrate that although non-Markovianity complicates quantum dynamics, its associated memory can be effectively exploited to boost the precision of parameter estimation in open quantum systems.

To further illustrate the effect of the extensions, Fig.~\ref{fig:histo} reports the NMSE at a fixed time step ($k=4$) for different values of $\lambda$. The bar heights represent the average NMSE, while the error bars indicate the standard deviation, representing the statistical variability across the 2000 independent realizations.
This representation highlights more clearly the comparative impact of the different strategies more clearly. For $\lambda=0.10$, the estimation remains challenging, although both temporal extensions lead to a visible reduction of the NMSE compared to the standard protocol, with the distant-step memory ($k^{\prime}=k_1$) performing best. At the intermediate value $\lambda=0.50$, all extensions improve the estimation, again with the fixed earlier step yielding the lowest error. In the Markovian limit ($\lambda=1.00$), the overall NMSE is significantly reduced, and the differences between strategies become smaller, yet the distant-step memory remains the most effective. These results confirm the dominant role of memory in enhancing estimation, with long-term memory providing the most robust improvements across different system frequencies.\\

\begin{figure}[ht]
    \centering
    \includegraphics[width=1.1\linewidth]{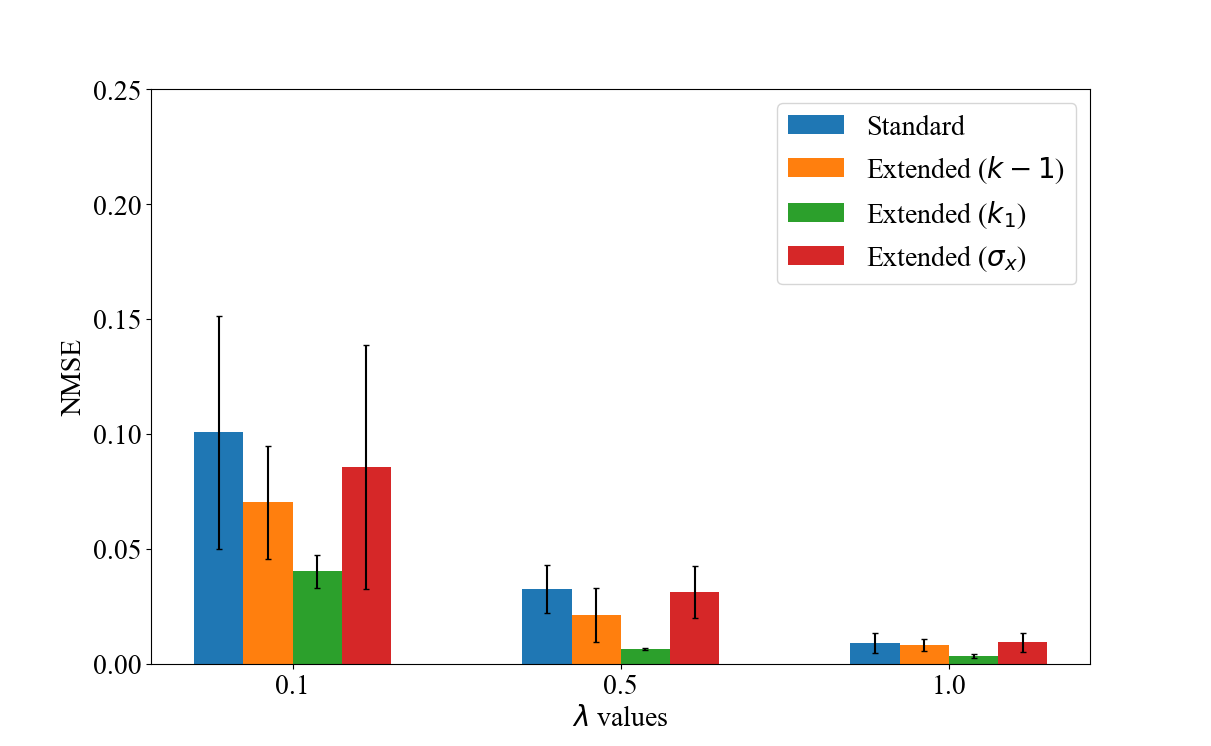}
    
    \caption{Comparison of the NMSE at time step $k=4$ for different depolarizing parameter $\lambda$, using the standard QELM (blue) and three extensions: immediate past $k^{'} = k -1$ (orange), distant step $k^{'} = k_1$ (green), and additional observable $\langle\sigma_x\rangle$ (red)(Average of 2000 realizations). }
    \label{fig:histo}
\end{figure}

\begin{figure}
    \centering
    \includegraphics[width=1\linewidth]{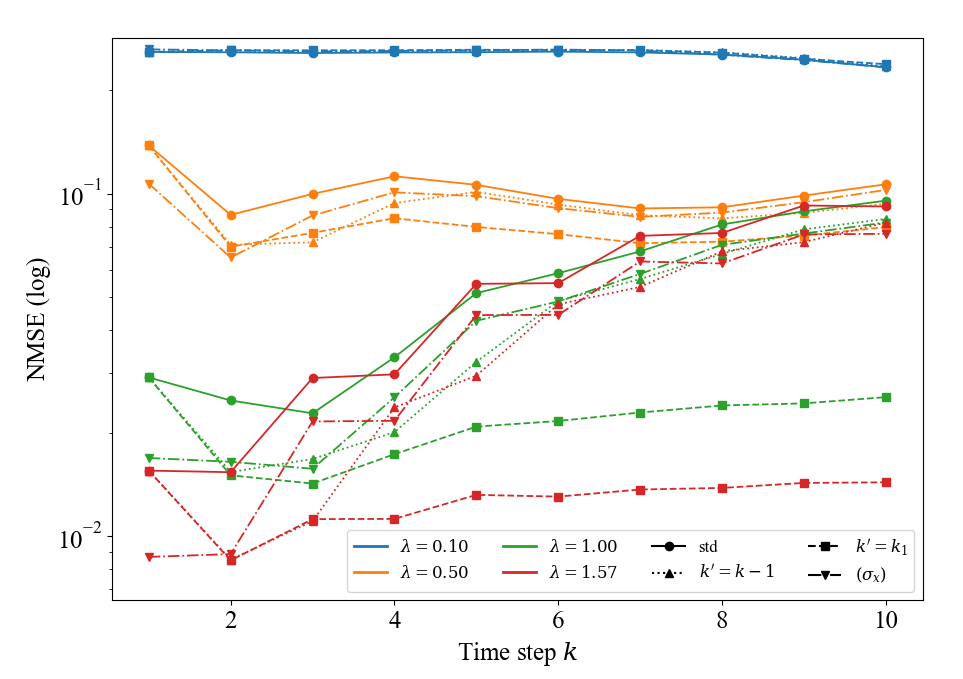}
    \caption{NMSE for the estimation of $\lambda$ with extended feature vectors. 
    Solid lines correspond to the baseline case, dotted and dashed lines represent memory extensions, and dash-dotted lines correspond to the use of $\langle\sigma_x\rangle$.}
    \label{fig:eta_ext}
\end{figure}

\textit{Parameter Estimation for $\lambda$.} In the second case, we fixed the parameter $\chi$ while $\lambda$ is varied. Fig.~\ref{fig:eta_ext} reports the NMSE as a function of time step for representative values of $\chi$, in the figure, Solid lines correspond to the baseline case, dotted and dashed lines represent memory extensions, and dash-dotted lines correspond to the use of $\langle\sigma_x\rangle$. We first focus on the baseline configuration. For weak coupling $(\chi = 0.1)$, the NMSE remains high and nearly constant, reflecting the difficulty of predicting the values of $\lambda$ with the QELM. By contrast, for stronger couplings $(\chi=0.5, 1.0, 1.57)$, the QELM achieves significantly lower errors at early time steps, although the NMSE gradually increases at longer times due to the reduced distinguishability of the system states. At $\chi=0.5$ the performance improves at short times and then stabilizes; at $\chi  = 1.00$ and $\chi = 1.57$, the NMSE achieves its minimum within the first few steps, reflecting efficient parameter encoding, but gradually increases at longer times.  
To improve the estimation further, we extended the feature vectors by incorporating memory and additional observables, so we have the vector~(\ref{x_k'}). Specifically, we tested three cases: (i) adding information from the immediate past step ($k' = k-1$), (ii) including features from a fixed earlier step ($k'=k_1$), and (iii) using $\langle\sigma_x\rangle$ as an additional observable. 

As shown in Fig.~\ref{fig:eta_ext}, solid lines correspond to the baseline case, while dotted and dashed curves illustrate the memory extensions, and the dash-dotted lines represent the use of $\langle\sigma_x\rangle$; all three strategies reduce the NMSE compared to the baseline. At $\chi=0.5$, the improvements are visible mainly at short times, while for $\chi=1.57$ the gains are stronger, with the memory extensions consistently outperforming the observable-based case.
These results demonstrate a clear hierarchy in information value for this task. While adding the $\sigma_x$ observable offers a marginal gain by enriching the instantaneous description of the reservoir state, the temporal extensions provide a qualitative improvement. They allow the linear readout to access the system's dynamical trajectory, which is crucial for disentangling the memory parameter $\lambda$ from the coupling strength $\chi$.

\section{Conclusion}
In this work, we investigated a quantum channel discrimination and parameter estimation protocol based on a QELM driven by the dynamics of a quantum collision model. By analyzing both Markovian and non-Markovian regimes, we demonstrated that the QELM can reliably learn and estimate key physical parameters, such as the system–environment coupling strength and the depolarizing rate, directly from the reduced system observables. The results show that the estimation accuracy depends strongly on the dynamical regime: nearly Markovian dynamics yield lower errors, whereas strongly non-Markovian conditions reduce the distinguishability of system trajectories.\\ 

The central contribution of this work is a systematic comparison of two strategies for enhancing the QELM feature space: (i) temporal memory extensions, where the feature vector at time $k$ is augmented with the reservoir state from an earlier time step $k'$ (immediate past $k-1$ or fixed reference $k_1$), and (ii) observable-based extensions, where we include measurements of an additional Pauli observable $\langle \sigma_x \rangle$ at the same time step. 

Our results reveal a key insight: temporal memory is a far more powerful resource than instantaneous observable diversity. While adding $\sigma_x$ provides only marginal gains, temporal extensions yield substantial and robust improvements in estimation accuracy. This indicates that for characterizing non-Markovian processes, the information encoded in the dynamical trajectory is more valuable than any single-time snapshot. 
Future directions include experimental implementation on quantum processors, exploration of alternative memory mechanisms, extension to simultaneous multi-parameter estimation, systematic study of optimal memory depth as a function of non-Markovianity, and application to other collision model architectures.

In summary, we have shown that a memory-augmented QELM can effectively characterize non-Markovian quantum dynamics. The comparison between temporal and observable-based extensions reveals that the key to unlocking information in complex open systems lies not just in measuring more, but in measuring across time.

\begin{acknowledgments}
This research was supported through computational resources of HPC-MARWAN (hpc.marwan.ma)provided by the
National Center for Scientific and Technical Research (CNRST), Rabat, Morocco. GLG acknowledges the Spanish State Research
Agency, through the María de Maeztu project CEX2021-001164-M, through the COQUSY project PID2022-140506NB-C21
and -C22, and through the QuantCom project CNS2024-154720,
all funded by MCIU/AEI/10.13039/501100011033. A.E.A. completed part of this work during a research visit to LPMMC, CNRS, in Grenoble, France. He extends his sincere gratitude to the CNRS - Fédération de Recherche QuantAlps, Comité de Direction QuantAlps, for their financial support and for fostering a stimulating and friendly research environment. 
  \end{acknowledgments}


\providecommand{\noopsort}[1]{}\providecommand{\singleletter}[1]{#1}%

\end{document}